\documentclass[aps,twocolumn,showpacs,preprintnumbers,prl,amsmath,
amssymb,amsfonts,superscriptaddress,floatfix]{revtex4}

\usepackage{amsfonts}
\usepackage{amssymb}
\usepackage{amsmath}
\usepackage{ulem}
\usepackage{color}
\usepackage{latexsym}
\usepackage{graphicx}
\usepackage{graphics}
\usepackage{floatflt}
\usepackage{epsfig}
\usepackage{overpic}
\usepackage{soul} 
\usepackage{varioref,xr-hyper}
\usepackage[latin1]{inputenc}

\newcommand{\be}{\begin{equation}}
\newcommand{\ee}{\end{equation}}
\newcommand{\bea}{\begin{eqnarray}}
\newcommand{\eea}{\end{eqnarray}}

\def\nn{\nonumber}
\def\lb{\label}
\def\pref#1{(\ref{#1})}

\def\ra{\rightarrow}

\def\bk{{\bf k}}

\def\bq{{\bf q}}

\def\bQ{{\bf Q}}

\def\d{\delta}

\def\l{\lambda}

\def\t{\tau}

\def\s{\sigma}
\def\o{\omega}
\def\G{\Gamma}
\def\D{\Delta}

\newcount\bozza \bozza=0
\ifnum\bozza=1
\newdimen\shift \shift=-2truecm
\def\lb#1{%
{\label{#1}\rlap{\kern\shift{$\scriptstyle#1$}}}}
\else\def\lb#1{\label{#1}} \fi

\begin{document}
\title{Orbital mismatch boosting nematic instability  in iron-based superconductors}
\author{Laura Fanfarillo}
\affiliation{CNR-IOM and International School for Advanced Studies (SISSA), Via
Bonomea 265, I-34136, Trieste, Italy}
\author{Lara Benfatto}
\email[corresponding author: ]{ lara.benfatto@roma1.infn.it}
\affiliation{ISC-CNR and Dep. of Physics, ``Sapienza'' University of Rome, P.le
A. Moro 5, 00185, Rome, Italy} 
\author{Bel\'en Valenzuela}
\email[corresponding author: ]{ belenv@icmm.csic.es}
\affiliation{Instituto de Ciencia de Materiales de Madrid, ICMM-CSIC,
Cantoblanco, E-28049 Madrid, Spain}
\date{\today}

\begin{abstract} 
{We derive the effective action for the collective spin modes in iron-based 
superconductors. We show that, due to the orbital-selective nature of spin 
fluctuations, the magnetic and nematic instabilities are controlled by the 
degrees of orbital nesting between electron and hole pockets. Within a 
prototypical three-pockets model the hole-electron orbital mismatch is found to 
boost spin-nematic order. This explains the enhancement of nematic order in FeSe 
as compared to 122 compounds, and its suppression  under pressure, where  
the emergence of the second hole pocket compensates the orbital mismatch of the 
three-pockets configuration.}
\end{abstract}

\pacs{74.70.Xa, 74.25.nd}
{\bf \maketitle }

Understanding the origin of the nematic phase is one of the most challenging 
open issues in the field of iron-based superconductors (IBS). In these systems the 
structural transition from tetragonal to orthorhombic is accompanied (and often 
preempted) by a marked electronic anisotropy which suggests an electronic origin 
of the instability \cite{GallaisReview16}. The original spin-nematic 
proposal \cite{schmalianprb12, fernandesnatphys14} focuses on the typical 
topology of the Fermi surface (FS) in pnictides, with hole-(h-)like pockets at 
$\Gamma$ and electron-(e-)like pockets at $\bQ_X=(\pi,0)$ and $\bQ_Y=(0,\pi)$ in 
the 1Fe unit-cell notation. The underlying idea is that the nesting between h- 
and e-pockets favors the spin fluctuations at these two equivalent momenta. 
According to \cite{schmalianprb12, fernandesnatphys14}, a nematic phase emerges 
since the ellipticity of the e-pockets induces an anisotropy of the paramagnetic 
spin fluctuations before that the long-range magnetic order sets in, lowering 
the symmetry of the electronic response from $C_4$ to $C_2$. This appealing 
scenario is  however challenged by the fact that nematicity is observed to be 
stronger or weaker in systems with similar band structure.

\begin{figure}[thb]
\begin{center}
\includegraphics[scale=0.45,clip=true]{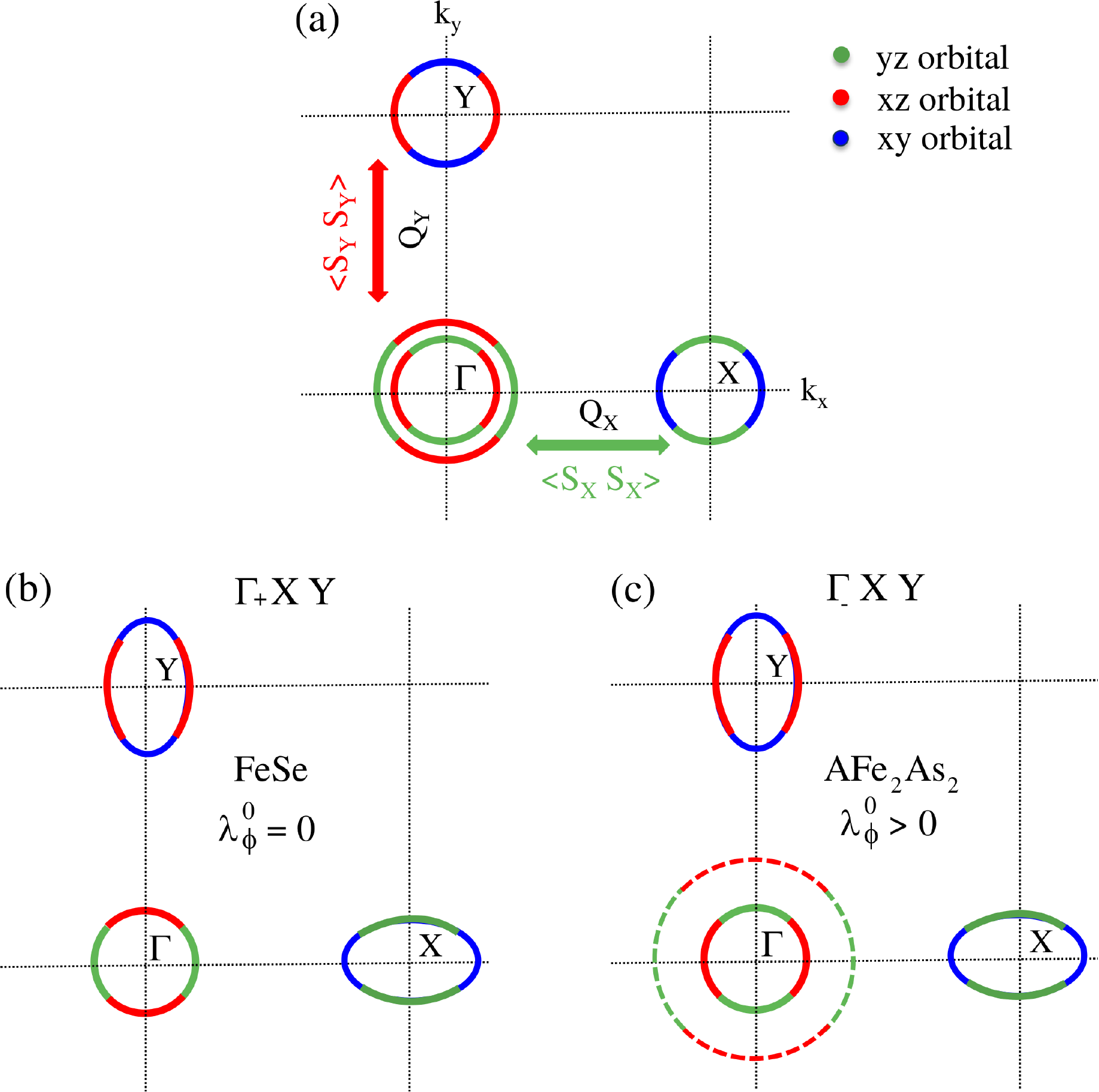} 
\caption{(a) General sketch of the orbital content of the FS of 
4-pocket model for IBS. The green/red arrows denote the OSSF, 
connecting h- and e-pockets at different momenta.  (b) Sketch of FeSe: only the 
outer pocket is present. (c) Sketch of 122 systems: the outer pocket is much 
larger, so it can be neglected in first approximation. The orbital mismatch 
(matching) in panel b (c)  is determined by the out-of-phase (in-phase) angular 
dependence of the $yz/xz$ orbital in the h- and $X/Y$ e-pockets.} 
\label{fig-OSSF}
\end{center}
\vspace{-1cm}
\end{figure}

FeSe is a remarkable example. Here the undoped compound has a
structural transition at $T_S=90K$ which is only cut-off below by the
superconducting transition at $T_c=9K$ \cite{BohmerPRB13}. The lack of magnetic
order motivated alternative interpretations for nematicity as due to orbital
ordering \cite{BaekNatMat14, SuJcondMat15, MukherjeePRL15, Jiangprb16,
Fanfarilloprb17, Chubukovprb17}. On the other hand, sizeable spin fluctuations
have been detected in FeSe as well \cite{Wang2016,Rahn15}, triggering an intense
investigation on the interplay between spin and orbital degrees of
freedom\cite{Fanfarilloprb15, Fernandesprb16, Fernandesreview17,
OnariKontaniprl16, GlasbrennerValentinatphys15, Fanfarillo2016, Chubukovprb17}.
Despite some interesting proposals \cite{Chubukovprx16, Kontaniprx16,
KontaniarXiv16}, no consensus has been reached yet on the mechanism favouring
nematicity in FeSe as compared to other systems, and leading to its suppression 
with external and internal pressure \cite{Sunnatcomm17, Kothapallinatcomm17,
Coldeareview17}.

In this Rapid Communication we show that the spin-nematic scenario is able to discriminate
topologically equivalent band structures once that the original derivation
\cite{schmalianprb12, fernandesnatphys14} is crucially revised accounting for
the orbital character of the bands. On general grounds, the
importance of the orbital content of the FS for the low-energy spin-fluctuations
in IBS, pointed out in \cite{Fanfarilloprb15}, has been recently discussed
within several contexts \cite{Khodasprb15, fernandesprb15, Fanfarillo2016,
Classenprl17}.
Here we show that the orbital topology of the FS crucially affects the
spin-nematic instability itself, which  is controlled by the degree of orbital
nesting, i.e. the relative orbital composition  between the h- and e-pockets
involved in the spin-exchange mechanism. By projecting the general microscopic
interaction\cite{dunghailee09, Aoki_prb09, graser09, Hirschfeldprb11,
Bascones16} on the low-energy multiorbital model of \cite{Vafekprb13} spin
fluctuations at different $\bQ$ vectors become orbital selective, i.e. they
involve only specific orbitals \cite{Fanfarilloprb15}, see Fig.~\ref{fig-OSSF}a. 
As a result also the interactions between spin modes beyond Gaussian level,
responsible for the nematic instability, becomes renormalized by the orbital
content of the h- and e-pockets. In particular we find that orbital nesting can
differentiate two topologically equivalent three-pocket models in which a single
hole pocket is present at $\Gamma$. In the case of FeSe the relevant h-pocket is
the outer one, see Fig.~\ref{fig-OSSF}b, and we find that its orbital mismatch
with the e-pockets boosts the nematic instability, while it is detrimental for
magnetism. In contrast, in the 122 family the most relevant h-pocket is the
inner one \cite{Finkprb15, Finkprb17, Avigoreview17}, having opposite orbital
character, see Fig.~\ref{fig-OSSF}b. In this case its good orbital nesting with
the e-pockets explains the robustness of the magnetic phase and the appearance
of a nematic instability only in its proximity.  Along the same reasoning, we
argue that in FeSe the suppression of nematicity  with internal or external
pressure \cite{Sunnatcomm17, Kothapallinatcomm17, Coldeareview17} can be
ascribed to the emergence of the inner hole pocket, changing the FS orbital
topology towards a more symmetric four-pocket model where nematicity can be easily
lost. 

We consider first a general four-pocket model with two h-pockets at $\G$, $\G_\pm$
and two e-pockets at $X$ and $Y$, that can be easily adapted to describe
different compounds among the 122 and 11 families. The kinetic part of the
Hamiltonian is derived adapting the low-energy model considered in
\cite{Vafekprb13}, where each pocket is described using a spinor
representation in the pseudo-orbital space \cite{Vafekprb13, Fanfarillo2016} 
\be
\lb{h0}
H_0^l=\sum_{\bk,\s} \psi^{l,\dagger}_{\bk\s} \hat H_0^l \psi^l_{\bk\s},
\ee
with $\hat H^l_0= h_0^l\t_0+\vec{h}^l\cdot\vec{\t}$, $l=\G,X,Y$ and $\t$ 
matrices represent the pseudo-orbital spin. The spinors are defined as 
$\psi^{\Gamma}=(c_{yz},c_{xz})$ and $\psi^{X/Y}=(c_{yz/xz},c_{xy})$. 
Diagonalizing $\hat H_0$ we find the dispersion relations $E^{l\pm}=h_0^l\pm 
h^l$ with $h^l=|\vec{h}^l|$. We introduce the rotation from the orbital to the 
band basis, 
\be
\lb{hole_fer}
\begin{pmatrix}
 h_+ \\
 h_- \\
\end{pmatrix} = 
\begin{pmatrix}
u_{\Gamma} &-v_{\Gamma}\\
v_{\Gamma} & u_{\Gamma}\\
\end{pmatrix}
\begin{pmatrix}
 c_{yz} \\
 c_{xz} \\
\end{pmatrix} 
\ee
with an analogous expression for the $X/Y$ pockets, provided that the 
corresponding orbital spinor is used. At  $X/Y$ only the $E^{X/Y+}$ band crosses 
the Fermi level, so in the following we will use $e_{X/Y}$ for the corresponding 
fermonic operators dropping the $+$ subscript.

% The interactions in the spin channel are described by the usual Hubbard $U$ and 
% Hund $J_H$ couplings, 
The interacting Hamiltonian is given by 
\be
\lb{hint}
H_{int}=-1/2\sum_{\bq \\'}U_{\eta\eta'} \vec{S}^\eta_{\bq } \cdot \vec{S}^{\eta'}_{-\bq}.
\ee
with $\eta,\eta'=yz,xz,xy$ denoting the orbital index. The interaction in the
spin channel is defined as $U_{\eta\eta'}\sim U \d_{\eta\eta'} + J_H
(1-\delta_{\eta\eta'})$, $U$ and $J_H$ being the usual Hubbard and Hund
couplings. We consider only spin operators with intraorbital
character $\vec{S}^\eta_{\bq }=\sum_{\bk ss'}(c^{\eta \dagger}_{\bk s} \vec \s
_{s s'}c^\eta_{\bk+\bq s'})$ with $\s_{ss'}$ are the Pauli matrices for the spin
operator. This choice is motivated by the general finding that intraorbital
magnetism is the dominant channel in IBS \cite{dunghailee09, Aoki_prb09,
graser09, Hirschfeldprb11, Bascones16}. The relevant magnetic fluctuations
occur at momenta $\bq$ near ${\bf Q}_X$ or ${\bf Q}_Y$. At low energy we can
project out the general interaction, Eq.~\pref{hint}, onto the fermionic
excitations defined by the model \pref{h0}. By using the rotation to the band
basis, Eq.~\pref{hole_fer}, one can then establish a precise correspondence
between the orbital and momentum character of the spin operators $\vec
S_{X/Y}^\eta\equiv \vec S_{\bq=\bQ_{X/Y}}^\eta$: 
\bea
\lb{sx}
\vec{S}_{X}^{yz} &=& \sum_\bk (u_{\G}h_{+}^{\dagger} 
+ v_{\G} h_{-}^{\dagger} )\, \vec{\s} \, u_{X} e_X \\
\lb{sy}
\vec{S}_{Y}^{xz} &=& \sum_\bk (-v_{\G}h_{+}^{\dagger} 
+ u_{\G} h_{-}^{\dagger} )\, \vec{\s} \, u_{Y} e_Y 
\eea
where we drop for simplicity the momentum and spin indices of the fermionic 
operators. It then follows that  the interacting Hamiltonian 
Eq.\pref{hint}  reduces to 
\be
\lb{hint_low}
H_{int}= - \frac{\tilde{U}}{2} \sum_{\bq \\'} \vec{S}^{yz/xz}_{X/Y} \cdot \vec{S}^{yz/xz}_{X/Y}.
\ee
where $\tilde{U}$ is the intraorbital interaction renormalized at low energy.
As it is clear from the above equation, it is the projection of the generic interaction Hamiltonian \pref{hint} onto the low-energy model \pref{h0} that  generates orbital-selective spin fluctuations (OSSF). Indeed, 
since at low energy the $xz/yz$-fermionic states exist only around
$Q_Y/Q_X$, it turns out that the spin operators $\vec{S}_{X}^{\eta}$ with $\eta\neq yz$ and $\vec{S}_{Y}^{\eta}$ with $\eta\neq xz$ are absent in Eq.\ \pref{hint_low}, so that there are no terms involving the Hund's
coupling. Once this correspondence has been establihed the derivation of the effective
action is formally equivalent to the one used in the simplified band language
\cite{schmalianprb12}. One can decouple the interaction term,
Eq.~\pref{hint}, by means of two vectorial Hubbard-Stratonovich (HS) fields 
$\vec{\D}_{X/Y}^{yz/xz}$ which will describe in what follows the collective
electronic spin fluctuations. The effective action up to quartic order becomes: 
\bea
\lb{Seff}
S_{\text{eff}} &=&
 \begin{pmatrix}
\D^{yz}_X & \D^{xz}_Y  \\
\end{pmatrix}
\begin{pmatrix}
\chi^{-1}_{X} & 0  \\
0 & \chi^{-1}_{Y}  \\
\end{pmatrix}
\begin{pmatrix}
\D^{yz}_X \\
\D^{xz}_Y  \\
\end{pmatrix}
\nn \\
&& \nn \\
&+& \begin{pmatrix}
(\D^{yz}_X)^2 & (\D^{xz}_Y)^2  \\
\end{pmatrix}
\begin{pmatrix}
u_{11} & u_{12}  \\
u_{12} & u_{22}  \\
\end{pmatrix}
\begin{pmatrix}
(\D^{yz}_X)^2 \\
(\D^{xz}_Y)^2  \\
\end{pmatrix}
\eea
Here $\chi^{-1}_{X/Y} = 1/U_s  + \Pi_{X/Y}^{yz/xz}$, where $U_s$ is the 
effective interactions between low-energy quasiparticles, and 
$\Pi^{yz/xz}_{X/Y}$ is the  propagator in the long-wavelength and zero-frequency 
limit:
\bea
\lb{piX_yz}
\Pi_{X}^{yz} &=& T\sum_{\bk,i\omega_n} u^2_{\G} u^2_{X} g_+ g_X  + v^2_{\G} u^2_{X} g_- g_X,   \\
\lb{piY_xz}
\Pi_{Y}^{xz} &=& T\sum_{\bk,i\omega_n} v^2_{\G} u^2_{Y} g_+ g_Y  + u^2_{\G} u^2_{Y} g_- g_Y.  
\eea
$g_{i} (\bk, i\o_n)= (i\o_n-E^{i}_\bk)^{-1}$ are the Green's functions 
in the band basis, $i=\pm$ denotes the h-bands and $i=X,Y$ the electronic 
ones. The coefficients of the quartic part of the action in Eq.\pref{Seff} are (see also \cite{fernandesprb15}):
\bea
\lb{u11}
u_{11}&=& T\sum_{\bk,i\omega_n}(u^2_X g_X)^2 (u^2_{\G} g_+ + v^2_{\G} g_-)^2,   \\
&& \nn \\
\lb{u22}
u_{22}&=&T\sum_{\bk,i\omega_n} (u^2_Y g_Y)^2 (v^2_{\G} g_+ + u^2_{\G} g_-)^2,   \\
&& \nn \\
\lb{u12}
u_{12}&=& T\sum_{\bk,i\omega_n} u^2_X g_X u^2_Y g_Y u^2_{\G} v^2_{\G}(g_+ - g_-)^2.
\eea
\begin{figure}[thb]
\begin{center} \includegraphics[scale=0.45,clip=true]{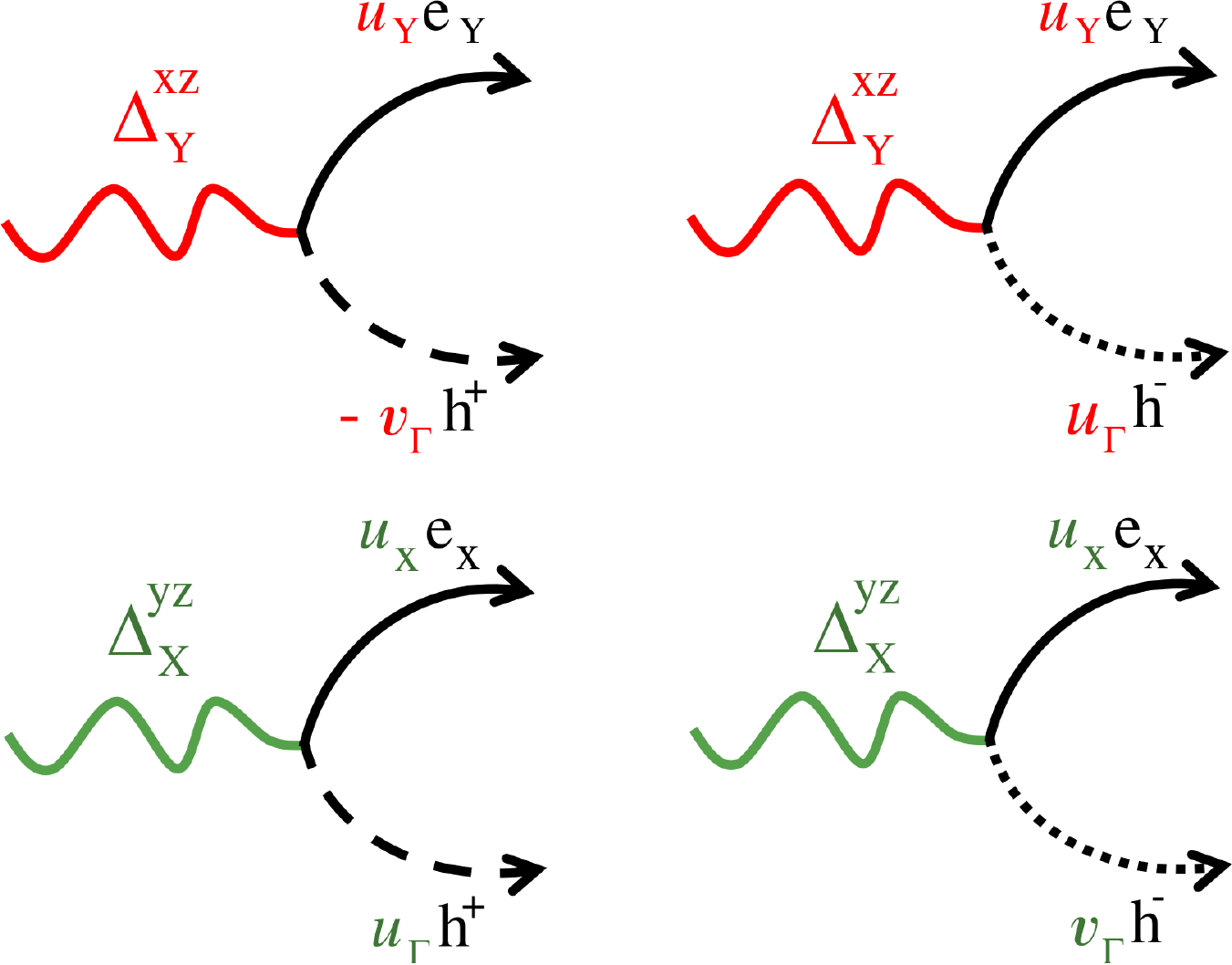} 
\caption{Diagrammatic representation of the vertices connecting the HS fields to 
the fermionic operators. Wavy (red/green) line denote the HS fields 
($\D_{X/Y}^{yz/xz}$), solid lines the excitations in the e-pockets, 
dashes/dotted lines excitations in the outer/inner h-pocket. The $u_l$, $v_l$ 
coefficients account for the orbital component of each band, according to the 
low-energy projection \pref{sx}-\pref{sy}.} \label{fig-Ver}
\end{center}
\vspace{-.8cm}
\end{figure}

As usual, the effective action is an expansion in powers of the 
HS fields. The coefficients of the $n^{th}$ power of the 
field is a loop with $n$ fermionic lines, leading to the product of two or four 
Green's functions in Eq.\ \pref{piX_yz}-\pref{piY_xz} and \pref{u11}-\pref{u12}, 
respectively. The vertices connecting $\Delta_{X/Y}^{yz/xz}$ to the band 
operators are depicted in Fig.\ \pref{fig-Ver}. Using this correspondence, which 
follows from the projection \pref{sx}-\pref{sy} of the spin operators at low 
energy, one easily understands that the fermionic loops are weighted with the 
elements $u_l,v_l$ defining the orbital content of each band. The magnetic 
instability is controlled by the Gaussian part of the action, Eq.~\pref{Seff}, 
and it occurs at the temperature where the inverse $Q_{X/Y}$ susceptibilities 
$\chi^{-1}_{X/Y} $ vanish. The nematic instability happens when the fluctuations 
along the $x$ and $y$ directions become inequivalent already above $T_N$. Since 
$u_{11}=u_{22}$ due to $C_4$ symmetry, the quartic part of the action, 
Eq.~\pref{Seff} can be simply diagonalized as 
\be
\lb{S4}
S^{(4)}_{\text{eff}} =  \lambda_\psi \psi^2 +\lambda_\phi \phi^2 
\ee 
where 
\bea
\lb{psi}
\psi &=& \frac{1}{\sqrt{2}} \big((\D^{yz}_{X})^2 + (\D^{xz}_{Y})\big)^2, \quad \lambda_\psi=u_{11}+u_{12}\\
\lb{phi}
\phi &=& \frac{1}{\sqrt{2}}\big( (\D^{yz}_{X})^2 - (\D^{xz}_{Y})\big)^2,  \quad \lambda_\phi=u_{11}-u_{12}.
\eea
%
% The above model has been used to address another topic, the spin reorientation due to 
% the spin-orbit coupling.\cite{fernandesprb15} 
% Concerning the simple form of the nematic order parameter, Eq.~\pref{phi}, 
% notice that the tensorial form of the nematic order parameter proposed in 
% Ref.~[\onlinecite{Fernandesprb16}] does not contain our result, 
% which in turn is dictated by the only possible non-Gaussian terms 
% Eq.~\pref{Seff} for the OSSF. 
Notice that the tensorial form of the nematic order parameter proposed in 
Ref.~[\onlinecite{Fernandesprb16}] does not contain our result Eq.~\pref{phi}, 
which in turn is dictated by the only possible non-Gaussian terms 
Eq.~\pref{Seff} for the OSSF.
From Eq.~\pref{S4} one sees that a nematic 
instability is possible only for $\lambda_\phi < 0$, when making $\langle 
\phi\rangle \neq 0$ lowers the energy of the system. However, while in 
Ref.~\cite{schmalianprb12} $\lambda_\phi$ is only controlled by the shape of the 
e-pockets, we find that also the degree of 
orbital nesting plays an important role. 

To make a first estimate of this effect we consider the simple case where 
the e-/h-pockets are perfectly nested circular FS, so that the orbital weights 
reduce to, $u_\G=u_Y=v_X=\cos \theta_\bk$, $v_\G=v_Y=u_X=\sin \theta_\bk$ 
% %
% \bea 
% \lb{uv}
% u_\G=u_Y=v_X=\cos \theta_\bk, \quad v_\G=v_Y=u_X=\sin \theta_\bk 
% \eea
% % 
and the Green's functions can be written as $g_X=g_Y=g_e=(i\omega_n-\epsilon)^{-1}$, 
$g_+=g_-=g_h=(i\omega_n+\epsilon)^{-1}$, with $\epsilon = -\epsilon_0 + \bk^2/2m  - 
\mu$. $\epsilon_0$ is the off-set energy, $m$ the parabolic band mass and 
$\mu$ the chemical potential. Within this approximation we can carry out 
explicitly the integration in Eq.s \pref{piX_yz}-\pref{u12}, showing that the 
differences between the various terms arise only from the angular integration of 
the product of the orbital weights. For what concerns the magnetic instability, 
the spin-fluctuations bubbles $\Pi_{X/Y}^{yz/xz}$, 
Eq.s~\pref{piX_yz}-\pref{piY_xz}, are both proportional to $\Pi_{eh} = 
T\sum_{\bk,i\omega_n} g_e g_h$ that lead to the usual log divergence: $\Pi_{eh} \sim
-N_F \log{\omega_0/T}$ where $N_F$ is the density of states and $\omega_0$ an 
upper cut-off \cite{supp-info}. On the other hand, the orbital renormalization 
of the $S^{(4)}_{\text{eff}}$ action is much more severe. Indeed, considering 
two hole pockets of same size, one immediately finds from Eq.\ \pref{u12} that 
$u_{12}=0$. This leads to a large {\it positive}  nematic eigenvalue 
$\lambda_\phi$ in Eq.~\pref{phi}, which prevents the occurrence of nematicity, 
in agreement with recent renormalization group studies on  the 4-pocket model 
\cite{Classenprl17}.

To simulate the case of specific compounds we consider two 3-pocket models in 
which a single hole pocket at $\Gamma$ is well-nested with the elliptical 
e-pockets: (a) The 3p$_+$ model for FeSe (Fig.~\ref{fig-OSSF}b), where only the 
outer pocket $\G_+$ crosses the Fermi level while the inner pocket $\G_-$ sinks 
below it before the nematic transition \cite{Fanfarillo2016, 
Coldeareview17}; (b) The $3p_-$ model for 122 systems (Fig.~\ref{fig-OSSF}c), 
where the outer pocket $\G_+$ is much larger than the electron ones, so it 
weakly contributes to the nesting \cite{ding08, hosonodaibook13}. These two 
models would be equivalent within the simplified band approach 
\cite{schmalianprb12} but lead to different OSSF actions. As far as nematicity 
is concerned, we see that  while the $u_{12}$ term in Eq.\ \pref{u12} is the 
same when only one of the two hole pockets is considered, the $u_{11}$ and 
$u_{22}$ terms pick up in a different way the orbital weights at $\Gamma$, 
allowing us to discriminate between  the two cases.  

{\it (a) FeSe:} As it has been recently discussed in \cite{Fanfarillo2016}, the 
disappearance of the inner hole pocket  in FeSe can be explained by the combined 
effect of spin-orbit coupling and OSSF shrinking mechanism. When only the $\G_+$ 
pocket is considered in Eq.s~\pref{u11}-\pref{u12} %one can easily see that 
all the coefficients of the quartic action become equal, so that at leading order 
$\lambda^0_\psi>0$ and $\lambda^0_\phi=0$.  Following the same lines of 
\cite{schmalianprb12}, we then include at perturbative level the e-pockets 
ellipticity and the deviations from perfect nesting. Since the results are 
robust with respect to the latter perturbation \cite{supp-info}, we discuss here 
only the dependence on the ellipticity parameter $\delta_e$. In this case, the 
eigenvalues of the quartic action turn out to be: 
\bea
\lambda_{\psi}^{3p_+}= 3  {\cal K} (T), \quad \quad
\lb{lfese}
\lambda_{\phi}^{3p_+}= - {\cal K}(T) \frac{b \, \delta^2_e }{T^2} 
\eea
with ${\cal K}(T)=7 N_F\zeta(3)/(8^3 \pi^2T^2)$. As one can see, as soon as a 
finite ellipticity is included, $\l_{\phi}<0$ at any temperature. This result is 
then analogous to the one found in the simplified band language of 
Ref.~\cite{schmalianprb12},  and the nematic critical temperature is determined 
by the divergence of the full nematic susceptibility $\chi_{nem}=\int_q 
\chi^2_{X}/(1+\l_\phi \int_q  \chi^2_{X})$ \cite{Fernandesreview12}.
On the other hand, the orbital mismatch between the h- and e-pockets realized 
in the case of FeSe is detrimental for the magnetic instability itself. Indeed,  
when only the $\G_+$ pocket is present the magnetic propagator in 
Eq.~\pref{piX_yz}-\pref{piY_xz} is reduced by a factor $1/8$ with respect to 
$\Pi_{eh}$ found in the simplified band language, since $\Pi_{X/Y}^{yz/xz} \sim 
\Pi_{eh} \int (d\theta/2\pi) \cos^2 \theta \sin^2\theta=\Pi_{eh}/8$ 
\cite{supp-info}. 

{\it (b) 122 systems:} In this case the good orbital nesting between the h- and 
e- pockets makes the $u_{11}$ term \pref{u11} much larger than the $u_{12}$ term 
\pref{u12}, so that at leading order $\lambda_\phi^0$ in Eq.\ \pref{phi} is {\it 
positive}, preventing a nematic transition. Accounting for the ellipticity of 
the e-pockets one finds: 
\be
\lambda_{\psi}^{3p_-}= {\cal K}(T) 
\bigg( 19  - \frac{12 \, b \, \delta^2_{e}}{T^2}    \bigg), \ \
\lambda_{\phi}^{3p_-}= {\cal K}(T) 
\bigg( 16  -  \frac{25}{2}  \frac{b\, \delta^2_{e}}{T^2}   \bigg),
\lb{l122}
\ee
so that the ellipticity is again the driving force for the nematic transition.  
However in this case  $\lambda_\phi^{3p_-}$ (which always becomes negative 
first) changes sign only below a temperature $T^{*}$ scaling as $T^* \sim 0.19 
\, \d_e $ \cite{supp-info}. At the same time the good orbital nesting pushes the 
magnetic transition to higher temperatures, since  $\Pi_{X/Y}^{yz/xz}\sim 
3\Pi_{eh}/8$.

\begin{figure}
\begin{center} \includegraphics[scale=0.68, clip=true]{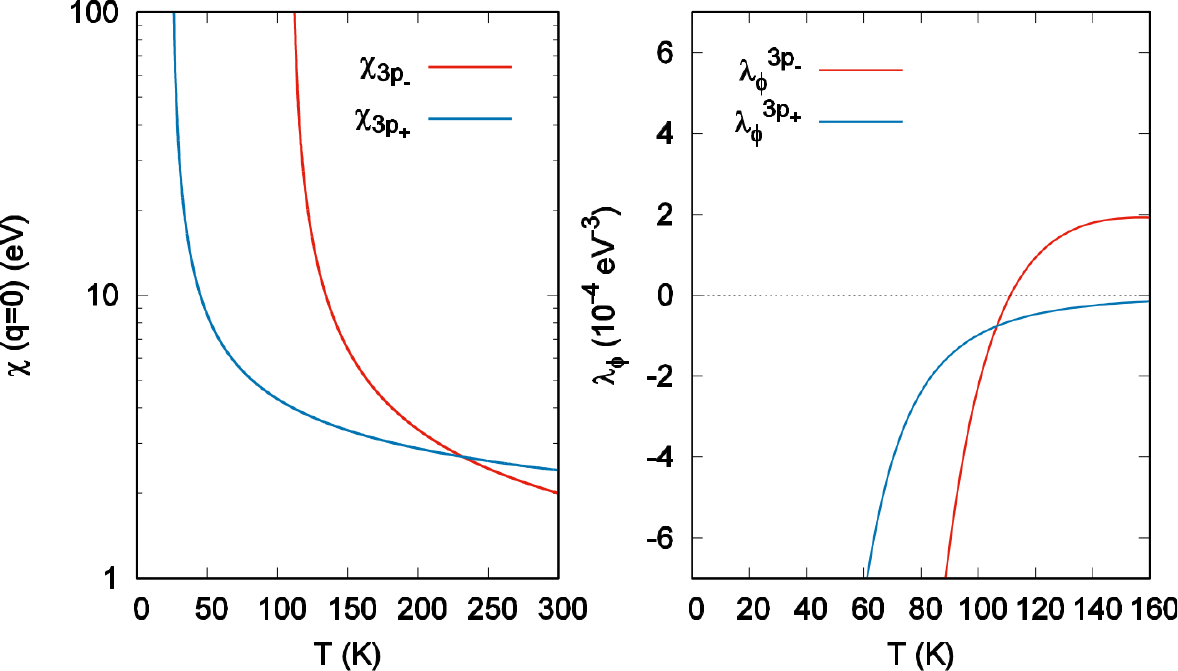}
\caption{(a)$\chi_{X/Y}^{yz/xz}$(q=0) and (b) nematic eigenvalue 
$\lambda_{\phi}$ for the 3p$_+$ and 3p$_-$ model for the same set of band 
parameters (see text).  Here $T_{Neel} = 110$, $24$ K for the  3p$_-$,  3p$_+$ 
model respectively, while the change of sign of $\lambda_{\phi}$ for the 3p$_-$ 
model occurs around $112$ K.} \label{fig:nem_coupling}
\end{center}
\vspace{-0.8cm}
\end{figure}

To make a quantitative comparison between the two 3pockets models we show in
Fig.~\ref{fig:nem_coupling} a-b the magnetic susceptibility
$\chi_{X/Y}^{yz/xz}$(q=0) and the nematic eigenvalue $\lambda_{\phi}$ using  the
same set of band parameters, as appropriate e.g. for 122 compounds
\cite{supp-info}. As one can see, by accounting uniquely for the different
orbital nesting the Ne\'el temperature of the 3p$_+$ model, $T_{Neel}^{3p_+}$,
is suppressed by about $80\%$ with respect to the 3p$_-$ case. Taking into
account also that the experimental density of states in FeSe is smaller than in
122 compounds \cite{Fanfarillo2016} $T_{Neel}^{3p_+}$ is expected to be further
suppressed \cite{supp-info}. Finally from Fig.~\ref{fig:nem_coupling}b, one
observes that while $\lambda_\phi^{3p_+}$ is always negative,
$\lambda_\phi^{3p_-}$ changes sign slightly above the $T^{3p_-}_{Neel}$, and
then rapidly increases in absolute value. These considerations provides a
possible explanation of the observed proximity between the nematic and magnetic
transition in 122 systems \cite{Indranilprb14}. 

The above results offers also a possible explanation for the suppression of
nematicity in FeSe under internal and external pressure. Indeed, it has been
reported that Sulphur isoelectronic substitution \cite{ColdeaarXiv17,
Coldeareview17} brings back the inner hole pocket above the Fermi level. This
finding is also supported by ab-initio calculations, which usually miss the
experimental position of the Fermi level but report in general an increase of
the hole-pockets size  with pressure \cite{KontaniarXiv16, roserprivate}. The
emergence of the inner hole pocket changes the FS topology of FeSe towards the
more symmetric 4-pocket model, which has been shown before to be detrimental for
nematicity, leading to the largest positive value of the $\lambda_\phi^0$
eigenvalue. On the other hand, the same mechanism could also
enhance magnetism, as observed. How these two effects interplay with the
concomitant increase of the superconductivity remains an open question for
future studies.

In conclusion, we derived the effective model for the spin fluctuations starting 
from a multiorbital low-energy 4-pockets fermionic model. We showed that orbital 
degrees of freedom renormalize the effective interactions between spin modes, 
with observable consequences on the magnetic and nematic instabilities. We 
considered explicitly a prototype 3-pockets model, as appropriate for FeSe and 
122 compounds, where the only difference between the two cases is the orbital 
content of the relevant h-pocket at $\Gamma$.  In FeSe the orbital mismatch 
between the outer h-pocket and the electron ones boosts nematicity and is 
detrimental for magnetism. In 122 compounds the good h-e orbital nesting 
favors magnetism and makes nematicity possible only at temperatures close to 
the magnetic transition. Our results offers a unified scenario to understand how 
orbital nesting can differentiate topologically equivalent band structures. 
Further confirmations of this mechanism can provide an useful tool to ultimately 
reach the external control on nematic order in iron-based systems.

\section*{Acknowledgment}

L.F. and B.V. acknowledges J\"org Fink for useful discussions. B.V. acknowledges Roser 
Valenti for discussion and for sharing her ab-initio calculations of FeSe with 
pressure. L.B. acknowledges financial support by Italian MIUR under project 
PRIN-RIDEIRON-2012X3YFZ2 and by MAECI under the Italian-India collaborative 
project SUPERTOP-PGR04879. B.V. acknowledges funding from MINECO (Spain) via Grants No.FIS2014-53219-P and 
Fundaci\'on Ram\'on Areces.

\bibliography{pnictides_nematic}

\section*{Supplementary Material for \\ 
Orbital mismatch boosting nematic instability  in iron-based superconductors}

In this supplemental material we discuss how to compute the quadratic and
quartic coefficients of the action and show the explicit results for the cases
of interest. We start from Eq.s (7)-(11) of the main text, that we rewrite here
for convenience:
\bea
\lb{piX_yz_app}
\Pi_{X}^{yz} &=&  T\sum_{\bk,i\omega_n} u^2_{\G} u^2_{X} g_+ g_X  + v^2_{\G} u^2_{X} g_- g_X,   \\
\lb{piY_xz_app}
\Pi_{Y}^{xz} &=&  T\sum_{\bk,i\omega_n} v^2_{\G} u^2_{Y} g_+ g_Y  + u^2_{\G} u^2_{Y} g_- g_Y.  
\eea
and 
\bea
\lb{u11_app}
u_{11}&=& T\sum_{\bk,i\omega_n}(u^2_X g_X)^2 (u^2_{\G} g_+ + v^2_{\G} g_-)^2,   \\
&& \nn \\
\lb{u22_app}
u_{22}&=&T\sum_{\bk,i\omega_n} (u^2_Y g_Y)^2 (v^2_{\G} g_+ + u^2_{\G} g_-)^2,   \\
&& \nn \\
\lb{u12_app}
u_{12}&=& T\sum_{\bk,i\omega_n} u^2_X g_X u^2_Y g_Y u^2_{\G} v^2_{\G}(g_+ - g_-)^2.
\eea
As already discussed the quartic part of the action can be simply 
diagonalized as $S^{(4)}_{\text{eff}} =  \lambda_\psi \psi^2 +\lambda_\phi 
\phi^2$ with
\bea
\lb{psi}
\psi &=& \frac{1}{\sqrt{2}} (\D^{yz}_{X})^2 + (\D^{xz}_{Y})^2, \quad \lambda_\psi=u_{11}+u_{12}\\
\lb{phi}
\phi &=& \frac{1}{\sqrt{2}} (\D^{yz}_{X})^2 - (\D^{xz}_{Y})^2,  \quad \lambda_\phi=u_{11}-u_{12},
\eea
where we used that $u_{11}=u_{22}$. An attractive nematic coupling
$\lambda_{\phi} <0 $ is a necessary (but not sufficient) condition for the
occurrence of the nematic transition, that happens only if a divergence of the
susceptibility $\chi_{nem}$ is found at a nematic critical temperature
$T_{nem}$. 

\section{Green Functions within the perfectly nested parabolic band approximation and beyond}

To make a first estimate of Eq.s~\pref{piX_yz_app}-\pref{u12_app} we consider
the simple case where the electron/hole pockets are perfectly nested circular
Fermi surfaces. The orbital weights simply become: 
\bea 
\lb{uv}
u_\G=u_Y=v_X=\cos \theta_\bk, \quad v_\G=v_Y=u_X=\sin \theta_\bk 
\eea
with $\theta_\bk = \arctan(k_y/k_x)$. One can also put 
$g_X=g_Y=g_e=(i\omega_n-\epsilon)^{-1}$ while 
$g_+=g_-=g_h=(i\omega_n+\epsilon)^{-1}$. Here $\epsilon$ is the parabolic 
dispersion $\epsilon = -\epsilon_0 + \bk^2/2m$, with $\epsilon_0$ is the off-set 
energy with respect to the chemical potential, put conventionally to zero, and 
$m$ the parabolic band mass.

As mentioned in the manuscript, we account for the deviations from the 
perfectly-nested parabolic-band approximation perturbatively. One can describe 
the ellipticity of the electronic band dispersion as
\be
\lb{de_per}
E^{X/Y}\simeq \epsilon \mp \delta_e \cos 2 \theta_\bk,  \quad \quad  
\d_e = \epsilon_0 \, m \, \big(\frac{m_x-m_y}{2m_x m_y} \big),
\ee
where $\delta_e$ accounts for the ellipticity of the electron pocket via the
$x/y$ anisotropy of the masses with respect the parabolic band mass $m$. The
expressions in Eq.s~\pref{de_per} correctly reproduces the opposite ellipticity
of the $X/Y$ pockets.
For the sake of completeness we also consider the deviation from perfect nesting 
due to e.g. mass, offset energy, spin-orbit coupling mismatch of the hole 
pockets via 
\be
\lb{dm_per}
E^{\Gamma_\pm}\simeq -\epsilon + \delta_{m_\pm},   \quad \quad 
\delta_{m_\pm}  = \epsilon_0 \bigg(\frac{m_\pm - m}{m} \bigg).
\ee
These perturbations can be included in the estimate of
Eqs.\pref{piX_yz_app}-\pref{u12_app} by expanding the Green functions for small
$\d_e$, $\d_{m_\pm}$: 
\bea 
\lb{per_gh}
g_\pm &=& g_h(1+\delta_{m_\pm}g_h) \\
\nn \\
\lb{per_ge}
g_{X/Y} &=& g_e(1\mp\delta_e \cos(2\theta_\bk)g_e).
\eea
In principle the perturbations $\d_e$ and $\d_{m_\pm}$ affect also the angular 
orbital factors, which should deviate from the $\cos \theta / \sin \theta$ 
expressions of Eq.\pref{uv}. However in first approximation we will neglect 
these modifications and we will retain only the effects of $\d_e$ and 
$\d_{m_\pm}$ on the Green's functions.

\section{Evaluation of the sum over frequency and momenta}

To compute the sum over Matsubara frequency and momenta in Eq.s\ 
\pref{piX_yz_app}-\pref{u12_app} we will use the usual decomposition: 
\be
\lb{spheric_coor}
\sum_{\bk} = \int_{BZ} \frac{d^2\bk}{(2 \pi)^2} = 
\int_0^{2\pi} \frac{d\theta}{2 \pi} \int d \epsilon N_F
\ee
where $\epsilon$ is the  energy, $\theta$ the azimuthal angle $N_F = m/2\pi$ is 
the density of state per spin at the Fermi level in 2D. In this way the only 
difference between the various models is in the angular integration of the 
orbital factors. Let us then discuss briefly the remaining  common integrals 
over energy and the Matsubara sums. 

Starting from the Gaussian term within the perfectly nested parabolic band approximation 
we need to compute the $\Pi_{eh}$ bubble
\be
\Pi_{eh} \equiv T N_F\sum_{i\omega_n}\int  d\epsilon \, g_e g_h 
\ee
By performing the energy integration via the calculus of the residua of the Green functions' poles 
we found 
\be
\Pi_{eh} = -  2 N_F T \sum_{n\geq0} \, \frac{\pi}{\, {\o_n}} = -  N_F \sum_{n\geq0} \, \frac{1}{(n+1/2)}
\ee
where we used that $ \omega_n = 2 \pi T (n+1/2)$. The calculation of the above sum can be 
carried out in terms of Euler digamma functions \cite{varlamovbook05}.
\be
\lb{digamma}
\psi^{(N)}(z) =  (-1)^{N+1}N!\sum^{\infty}_{n=0}\frac{1}{(n+z)^{N+1}} .
\ee 
The logarithmic divergence at the upper limit ($\psi^{(0)}(z>>1) \sim ln(z)$) is 
cut-off by the $\omega_0$ typical energy scale of the spin mode and one gets
\bea
\Pi_{eh} &=& -  N_F \bigg(\psi^{(0)} \bigg(\frac{1}{2} + \frac{\omega_0}{2 \pi T}\bigg) 
- \psi^{(0)}\bigg(\frac{1}{2}\bigg)\bigg) \nn \\
&=& -N_F \bigg( ln (\o_0/T) + ln (2/\pi) + C_E\bigg)
\lb{log}
\eea
where we used that $\psi^{(0)}(1/2) = - C_E - 2 ln(2)$ with $C_E$ being the 
Euler-Mascheroni constant. 

In order to compute the quartic terms, Eq.s~\pref{u11_app}-\pref{u12_app}, 
within the perfectly nested parabolic band approximation we need to compute
\be
\lb{res3}
T\sum_{i\omega_n}\int d\epsilon  \, g_e^2 g_h^2 = 
\,  T \sum_{n\geq 0} \, \frac{\pi}{{\o_n}^3} \\ 
\ee
while beyond such approximation the Green functions expansion lead to:
\bea
T\sum_{i\omega_n}\int d\epsilon  \, g_e^4 g_h^2 &=& 
-  T \sum_{n\geq0} \, \frac{\pi}{2 \, {\o_n}^5} \nn \\
T\sum_{i\omega_n}\int d\epsilon \, g_e^3 g_h^3 &=& 
-  T \sum_{n\geq0} \, \frac{3 \, \pi}{4\,  {\o_n}^5} 
\lb{res5}
\eea
and analogously for the $g_e^2 g_h^4$ case. 
It is easy to verify that the integrals of combination $(g_eg_h)^{m_1} 
g_{e/h}^{2m_2+1}$ with odd unpaired powers of the e/h Green's functions vanish, 
since the contribution coming from Matsubara frequency with positive $n$ exactly 
cancels out with the contribution of the negative ones. Using that $ \omega_n = 
2 \pi T (n+1/2)$, one can recognize in Eq.s~\pref{res3}-\pref{res5} the Euler 
digamma functions, Eq.\pref{digamma} for $z=1/2$ and $N=2, 4$  
\cite{varlamovbook05}. For $z=1/2$ one can express $\psi^{(N)}(1/2)$ in terms of 
the Riemann $\zeta(n)$ functions as 
$$ \psi^{(N)}(1/2) = (-1)^{N+1}N! \, (2^{N+1} -1) \zeta(N+1) $$
Using these definitions in Eq.s~\pref{res3}-\pref{res5} we obtain
\bea
\lb{AT}
T\sum_{i\omega_n}\int d\epsilon  \, g_e^2 g_h^2 &=&
\phantom{-}
\, \frac{7  \zeta(3)}{8\pi^2T^2} \equiv \, {\cal A}(T)\\ 
\lb{BT}
T\sum_{i\omega_n}\int d\epsilon \, g_e^4 g_h^2&=&
-\frac{31  \zeta(5)}{64\pi^4\,T^4} \equiv  \, {\cal B}(T) \\
\lb{CT}
T\sum_{i\omega_n}\int d\epsilon \, g_e^3 g_h^3&=&
- \frac{93 \,  \zeta(5)}{128\pi^4\,T^4}
\equiv  \, {\cal C}(T)  
\eea
where $\zeta(3)\sim 1.202$ and $\zeta(5)\sim1.037$, from which it follows that 
\be
\lb{BCA}
{\cal B}(T) = -\frac{ b \, {\cal A}(T)}{T^2} \quad \quad 
{\cal C}(T) = - \frac{3 \, b {\cal A}(T)}{2\,T^2}
\ee
with $b\sim 0.048$.

\section{ESTIMATE OF THE QUADRATIC AND QUARTIC TERMS OF THE ACTION} 

\subsection{3-pocket model $\Gamma_+ X Y$: complete orbital mismatch}

In this case we need to account only for the contribution coming from the 
$\Gamma_+$ in Eq.s \pref{piX_yz_app}-\pref{u12_app}. The quadratic term in the 
$q=0$ limit is given by
\bea
\lb{pi_3p}
\Pi_{X/Y}^{yz/xz} &= & T \sum_{\bk,i\omega_n} g_h g_e  \sin^2 \cos^2 \theta = \frac{\Pi_{eh}}{8} \nn \\
&=& - \frac{ N_F}{8} \bigg(\ln(\omega_0/T) + const\bigg)
\eea
with $const = ln (2/\pi) + C_E$. Here we separated the integrations as in 
Eq.\pref{spheric_coor}, used the results of Eq.\pref{log} and performed the 
angular integral $\int (d\theta/2\pi)\sin^2 \theta \cos^2 \theta = 1/8$. The 
Ne\'el temperature is determined as the temperature at which the pole of the 
magnetic susceptibility occurs 
\be
\lb{chi3p+}
{\chi^{{yz/xz}}_{X/Y}}^{-1}(q=0) = \frac{1}{U_s} +\Pi_{X/Y}^{yz/xz} 
= \frac{ N_F }{8} \ln \bigg( \frac{T}{T_{Neel}}\bigg)
\ee
where
\be
\lb{TNeel3p+}
T_{Neel} =  1.13\, \omega_0 \, e^{- 8/(N_F \, U_S)}.
\ee

Concerning the quartic term, we have that within the perfectly-nested 
parabolic-band approximation  all the quartic coefficients, 
Eq.s~\pref{u11_app}-\pref{u12_app}, are equivalent $u^0_{11/22}=\, u^0_{12} = 
u^0$ 
\bea
u^0 = T \sum_{\bk,i\omega_n} g_h^2\,  g_e^2 \sin^4 \theta \cos^4 \theta 
= \frac{3 \, N_F}{128} {\cal A}(T)
\eea
where we borrowed the results from Eq.~\pref{AT} and computed the angular
integral. In this case from the diagonalization of the quartic form,
Eq.s~\pref{psi}-\pref{phi}, we obtain 
\be
\lambda^0_{\psi}=\frac{3 \, N_F}{64} {\cal A}(T) \quad \quad  \lambda^0_{\phi}=0
\ee

Beyond the perfectly-nested parabolic-band approximation we can include the 
effects of the band nesting mismatch of the $\Gamma_+$ pocket and of the 
ellipticity of the electron pocket using the Green functions' expansion of 
Eq.s~\pref{per_gh}-\pref{per_ge}. With simple steps by using the integrals 
Eq.s~\pref{AT}-\pref{BT} one can easily obtain the expressions for the $u_{ij}$ 
terms up to order $\d_e^2$, $\d_{m_\pm}^2$
\be
u_{11/12} = \frac{ N_F \, {\cal A}(T)}{128} 
\bigg[ 3  - \frac{b}{T^2} \bigg(3 \delta^2_{m_+} \pm \frac{1}{2} \delta^2_{e} \bigg)\bigg] 
\ee
where we further simplified our expressions accounting for the relation between 
${\cal A}(T)$ and ${\cal B}(T)$ (see Eq.~\pref{BCA}). It is now straightforward 
to compute the $\lambda_{\psi/\phi}$ coupling
\bea
\lb{lpsifese}
\lambda_{\psi}&=& \frac{3\,  N_F \,  {\cal A}(T)}{64} \bigg( 1  - \frac{b}{T^2} \delta^2_{m_+} \bigg) \\
\lb{lphifese}
\lambda_{\phi}&=& - \frac{ N_F \, {\cal A}(T)}{64} \frac{b}{T^2} \delta^2_e 
\eea
as quoted in the main manuscript, with the definition ${\cal K}(T)=N_F{\cal A}(T)/64$.

\subsection{3-pocket model $\Gamma_- X Y$: the perfect orbital match}

We proceed in analogous way to compute the quadratic and quartic coefficients 
for the other cases of interest. Within a 3-pocket model $\Gamma_- X Y$ we 
account for the contribution of the $\Gamma_{-}$ pocket only. The quadratic term 
in the $q=0$ limit in this case is given by
\bea
\lb{pi_3m}
\Pi_{X}^{yz} &=& T \sum_{\bk, i\omega_n} g_h g_e \sin^4 \theta =  \frac{3}{8}\Pi_{eh}\nn \\
&=& - \frac{3 N_F}{8} \bigg( \ln(\omega_0/T)  + const\bigg)
\eea
where we used  $\int (d\theta/2\pi)\sin^4 \theta = 3/8$. As expected the same 
result is found for $\Pi_{Y}^{xz}$ where the angular factor goes like $\int 
(d\theta/2\pi)\cos^4 \theta = 3/8$. The magnetic susceptibility is given by 
\be
\lb{chi3p-}
{\chi^{{yz/xz}}_{X/Y}}^{-1}(q=0) = \frac{3 N_F }{8} \ln \bigg( \frac{T}{T_{Neel}}\bigg)
\ee
and the Ne\'el temperature is 
\be
\lb{TNeel3p-}
T_{Neel} = 1.13\, \omega_0 \, e^{- 8/(3 \, N_F \, U_S)}.
\ee
Concerning the quartic coefficients, within the perfectly-nested parabolic-band 
approximation, we have
\bea
&u^0_{11/22} &= T \sum_{\bk,i\omega_n} g_h^2\,  g_e^2 \sin^4 \theta \sin^4 \theta 
= \frac{35 \, N_F}{128} {\cal A}(T)\nn \\
&u^0_{12}& = T \sum_{\bk,i\omega_n} g_h^2\,  g_e^2 \sin^4 \theta \cos^4 \theta 
= \frac{3 \, N_F}{128} {\cal A}(T)
\eea
Since now $u_{11/22} \neq u_{12}$ the diagonalization of the quartic form
Eq.s~\pref{psi}-\pref{phi} lead to a finite $\lambda^0_{\phi}$ 
\be
\lambda^0_{\psi}=\frac{19 \, N_F}{64} {\cal A}(T) \quad \quad  
\lambda^0_{\phi}=\frac{16 \, N_F}{64} {\cal A}(T)
\ee

The contributions coming from the next orders $\d_e$, $\d_{m_-}$ can be computed
following the same approach used in the previous section.  Through tedious but
straightforward calculations, using Eq.s~\pref{AT}-\pref{CT} for computing
the integrals and the relations among ${\cal A}(T)$, ${\cal B}(T)$ and ${\cal
C}(T)$ of Eq.~\pref{BCA}, one arrives at 
\bea
\lambda_{\psi}&=& \frac{N_F \,  {\cal A}(T)}{128} 
\bigg( 38  - \frac{b}{T^2} \big( 38 \delta^2_{m_-} 
+ 24 \delta^2_{e} + 168 \delta_{e} \delta_{m_-} \big) \bigg) \nn \\
\lambda_{\phi}&=& \frac{N_F \, {\cal A}(T)}{128} 
\bigg( 32  - \frac{b}{T^2} \big( 32 \delta^2_{m_-} 
+ 25 \delta^2_{e} + 168 \delta_{e} \delta_{m_-} \big) \bigg). \nn \\
\lb{l122}
\eea
that in the limit $\d_{m_-}\ra 0 $ reduce to the expressions quoted in the main manuscript.

\subsection{4-pocket: $\G_{\pm}, X, Y$}

Within the 4-pocket model both the outer and inner hole pockets contribute to
the quadratic and quartic coefficients of the action
Eq.s~\pref{piX_yz_app}-\pref{u12_app}. 
The quadratic term in the $q=0$ limit in this case is given by
\bea
\lb{pi_4p}
\Pi_{X}^{yz} &=&  T \sum_{\bk, i\omega_n} g_h g_e  (\cos^2 \theta \sin^2 \theta + \sin^4 \theta) 
=  \frac{\Pi_{eh}}{2}  \nn\\
&=&- \frac{N_F}{2} \bigg( \ln(\omega_0/T)    + const\bigg) 
\eea
and analogous for $\Pi_{Y}^{xz}$. The magnetic susceptibility is
given by 
\be
\lb{chi4p}
{\chi^{{yz/xz}}_{X/Y}}^{-1}(q=0) = \frac{N_F}{2} \ln \bigg( \frac{T}{T_{Neel}}\bigg)
\ee
where the Ne\'el temperature is 
\be
\lb{TNeel4p}
T_{Neel} = 1.13\, \omega_0 \, e^{- 2/(N_F \, U_S)}.
\ee
Within the perfectly-nested parabolic-band approximation the quartic coefficients
go as
\bea
&u^0_{11} &= T \sum_{\bk,i\omega_n} g_h^2\,  g_e^2 \sin^4 \theta 
= \frac{3 \, N_F}{8} {\cal A}(T)\nn \\
&u^0_{22} &= T \sum_{\bk,i\omega_n} g_h^2\,  g_e^2 \cos^4 \theta 
= \frac{3 \, N_F}{8} {\cal A}(T)\nn \\
&u^0_{12}& = T \sum_{\bk,i\omega_n} g_e^2 (g_h -g_h)^2 \sin^4 \theta \cos^4 \theta = 0
\eea
Since here $u_{12}=0$ from Eq.s~\pref{psi}-\pref{phi} we find two identical coupling 
at the lower order  
\be
\lambda^0_{\psi}=\lambda^0_{\phi}=\frac{3 \, N_F}{8} {\cal A}(T)
\ee

The effect of the perturbations $\d_e$, $\d_{m_-}$ can be computed as before and 
contributes to the $\lambda_{\psi/\phi}$ couplings as 
\bea
\lambda_{\psi}&=& \frac{ N_F \, {\cal A}(T)}{64} 
\bigg( 24  - \frac{b}{T^2} \big( 19 \delta^2_{m_-} 
+ 14 \delta^2_{e} + 90 \delta_{e} \delta_{m_-} \big) \bigg) \nn \\
\lambda_{\phi}&=& \frac{N_F \, {\cal A}(T)}{64} 
\bigg( 24  - \frac{b}{T^2} \big( 16 \delta^2_{m_-} 
+ 14 \delta^2_{e} + 90 \delta_{e} \delta_{m_-} \big) \bigg). \nn \\
\lb{l4p}
\eea

\section{Quantitative Analysis of the results for the 3-pocket, $\G_+ X Y $,
$\G_- X Y $, and 4-pocket, $\G_\pm X Y $, models}

To elucidate the effects of the orbital mismatch on suppressing magnetism and 
boosting nematicity, we will consider band-structure parameters appropriate for 
122 iron-based compounds e.g. BaFe$_2$As$_2$. For the spin fluctuations we refer 
to \cite{Inosov10} and use $\omega_0 \sim 18$ meV. \\

We first consider the difference in Ne\'el temperatures, $T_{Neel}$, for the two 
the 3-pocket models, $\G_+ X Y $ (3p$_+$) and $\Gamma_- X Y$ (3p$_-$). To 
determine $T_{Neel}$ we need the value of the low-energy coupling $U_s$. We 
choose this value in order to reproduce, within the $\G_- X Y $ model, the 
experimental value $T_{Neel}\sim 110$ K found for weakly doped BaFe$_2$As$_2$ 
compounds \cite{Paglione10}. Keeping then all the parameters fixed we can 
estimate the value of $T_{Neel}$ in the  $\G_+ X Y $ model, which only differs 
in the orbital composition of the hole pocket at $\Gamma$. From  
Eq.~\pref{TNeel3p+} we then obtain $T_{Neel} = 24$ K, i.e. a suppression of the 
$\sim 80\%$ with respect the $\G_- X Y $ model, uniquely due to the different 
degree of orbital matching between hole and electron pockets of the two cases. A 
more precise estimate for FeSe would require to account also for the different 
band-structure parameters in the two cases. In particular  FeSe is characterized 
by electron pockets with a density of states $N_F$ about 30$\%$ smaller than in 
122 compounds \cite{Fanfarillo2016}. If we account for this difference in 
Eq.~\pref{TNeel3p+} the $T_{Neel}$ of the $\G_+ X Y $ model, used to describe 
FeSe, is further suppressed, approaching the critical temperature of the 
superconducting instability of FeSe. 

For sake of completeness we also compute $T_{Neel}$ for the 4-pocket model from 
Eq.~\pref{TNeel4p}. In this case, since both the inner and outer pockets 
$\Gamma_{\pm}$ contribute to the instability, the Ne\'el temperature reaches the 
$130$ K.

\begin{table}[!h]
\begin{tabular}{c c c c}
\hline
\phantom{XX} Model: \phantom{XX}	 & \phantom{X} 3p$_+$ (FeSe) \phantom{X} 
&  \phantom{X}3p$_-$ \phantom{X} & 4p  \phantom{X}\\
\hline
\hline
\phantom{XX}\phantom{XX}  & \phantom{XXXX} \phantom{X} \\
\phantom{XX}$T_{Neel}$ (K)\phantom{XX}	 &  24  & 110 & 132 \\
\phantom{XX}\phantom{XX}  & \phantom{XXXX} \phantom{X} \\
\hline
\end{tabular}
\caption{$T_{Neel}$ for the 3-pocket models, $\Gamma_+ X Y$ (3p$_+$) and 
$\Gamma_- X Y$ (3p$_-$), and for the 4-pocket model $\Gamma_\pm X Y$ (4p). 
Assuming $T_{Neel} \sim 110$ K for the 3p$_-$ model we estimate $N_F U_s$ and 
use this values to compute the Neel temperature of the other cases. For the 
3p$_+$ model we find a suppression of the $80\% $  with respect the 3p$_-$ case 
due to the orbital mismatch. In FeSe we expect an even stronger suppression 
since its experimental density of states is lower than in the 122 family. A 
higher $T_{Neel}$ is found instead in the 4p model, in which both the inner and 
outer hole pockets contribute to the magnetic instability.} 
\label{table-Neel} 
\end{table}

Once computed the Ne\'el temperature for the various cases (TABLE: 
\ref{table-Neel}), we can easily compute the $q=0$  magnetic susceptibility as 
in Eq.s~\pref{chi3p+}, \pref{chi3p-}, \pref{chi4p}. We show in Fig.~3a of the 
main text the temperature dependence of the susceptibility $\chi = 
\chi^{yz/xz}_{X/Y}$ for the 3p$_-$ /3p$_+$ models using $T_{Neel}=110$/$24$ K 
respectively. The density of states $N_F\sim 1.3$ eV$^{-1}$ is derived assuming 
$1/(2m)\sim 60$ meV. The typical logarithmical divergence is found at the 
correspondent $T_{Neel}$. It is interesting to notice that due to the different 
numerical prefactors ($1/8$ vs $3/8$) in Eq.s~\pref{chi3p+} and \pref{chi3p-} 
$\chi(q=0)$ takes similar value for the two model around room temperature even 
if the divergence of the $\chi_{3p_+}$ is found at lower temperature with 
respect to the 3p$_-$ case.\\

Concerning the quartic-order coefficients of the action, a qualitative analysis of 
Eq.s~ \pref{lpsifese}-\pref{lphifese}, \pref{l122} and \pref{l4p} shows that: 
\begin{itemize}
 \item[(i)]For the 3p$_+$ model (FeSe case) independently on others parameters
the nematic coupling $\lambda_\phi$  Eq.~\pref{lphifese} is negative as soon as
the ellipticity of the electron pockets is considered, while the symmetric
eigenvalue $\lambda_\psi$ Eq.~\pref{lpsifese} remains finite and positive. 
\item[(ii)]For the 3p$_-$ model, as one can check from Eq.~\pref{l122}, at any
value of $\delta_{m_-}$ the first eigenvalue which becomes negative for
decreasing temperature is $\lambda_\phi$. Thus the ellipticity is again the
driving force for the nematic transition. However here we need a finite value of
$\d_e$ in order to induce a sign change in $\lambda_\phi$. Putting
$\delta_{m_-}=0$ one can derive the temperature $T^{*}$  below which the nematic
eigenvalue becomes negative as a function $\d_e$, 
$T^{*} = ({25 \,b/32})^{1/2}  \d_e \ \  \sim \ 0.19 \ \d_e$
\item[(iii)]For the complete 4p model a nematic instability is prevented by the 
sign of the nematic eigenvalue (see Eq.~\pref{l4p}). Indeed in this case at any 
finite value of $\delta_{m_-}$ the first eigenvalue which becomes negative for 
decreasing temperature is $\lambda_\psi$. Assuming $\delta_{m_-}=0$ the 
correction to $\lambda_\psi$ and $\lambda_\phi$ reduces to the identical 
$\delta_e^2$ term thus the temperature $T^{*}$ below which the nematic 
eigenvalue becomes negative is, as a matter of fact, the same that determines 
the change of sign of $\lambda_\psi$.
\end{itemize}

\begin{table}
\begin{tabular}{c c}
\hline
\multicolumn{2}{c}{\phantom{XXXXXX}BAND Parameters\phantom{XXXXXX}} \\
\hline
\hline
\phantom{XX}\phantom{XX}&\phantom{XXXX}\phantom{X} \\
\phantom{XX}$\epsilon_0$ (meV)\phantom{XX}	& \phantom{XXXX} 90 \phantom{X} \\
\phantom{XX}$1/(2m)$ (meV)\phantom{XX}		& \phantom{XXXX} 60 \phantom{X} \\
\phantom{XX}$N_F$  (ev$^{-1}$)\phantom{XX}	& \phantom{XXXX} 1.3 \phantom{X}\\
\phantom{XX}$k_F^0$ ($\pi/a$)\phantom{XX}	& \phantom{XXXX} 0.31 \phantom{X} \\
\phantom{XX}$\d_e/\epsilon_0$  \phantom{XX}	& \phantom{XXXX} 0.55 \phantom{X} \\
\phantom{XX}${k_F}_x $ ($\pi/a$)\phantom{XX}	& \phantom{XXXX} 0.39 \phantom{X} \\
\phantom{XX}${k_F}_y$ ($\pi/a$)\phantom{XX}	& \phantom{XXXX} 0.21 \phantom{X} \\
\phantom{XX}\phantom{XX}  & \phantom{XXXX} \phantom{X} \\
\hline
\end{tabular}
\caption{Band parameters appropriate for 122 compounds used in this section.}
\label{table-par}
\end{table}

Finally, we need to choose some parameter values for a quantitative estimate of 
the nematic coupling of the 3 pocket models used in Fig.~3b of the main text. 
For the band structure we choose again parameters appropriate for weakly-doped 
122 compounds:  we set $\epsilon_0 = 90 $ meV and $1/(2m) \sim 60$ meV, ($N_F 
\sim 1.3$ eV$^{-1}$). With these parameters we have circular Fermi pockets of 
radius $k_F^0 \sim 0.31$ in $\pi/a$ unit, $a\sim 3.96~\text{\AA}$ is the lattice 
parameter. Beyond the parabolic-band approximation we further consider the 
ellipticity of the electron pockets Eq.~\pref{de_per}  assuming $\d_e = 0.55 
\epsilon_0$. This define electronic elliptical Fermi pockets with $k_F^{x/y} 
\sim 0.39$ and $k_F^{y/x} \sim 0.21$ for the X/Y pockets respectively. For 
simplicity we consider the case $\d_{m_-}=0$. Any finite value of such a 
perturbation lead to similar conclusions. All the band parameters used are 
collected in TABLE:~\ref{table-par}. The nematic eigenvalue $\lambda_{\phi}$ for 
the 3p$_+$ model Eq.~\pref{lphifese} and for the 3p$_-$ model Eq.~\pref{l122} as 
a function of $T$ are plotted in Fig.~3b in the main text. While for the 3p$_+$ 
case the nematic coupling is negative at any temperature, for the 3p$_-$ model 
we need to cool the system below $T^* \sim 112$ K in order induce a sign change 
in \ the nematic eigenvalue. Below this temperature the absolute value of 
$\lambda_\phi$  grows rapidly with decreasing $T$, explaining why in this 
system, where $T_{Neel} = 110$ K, the nematic transition occurs very close to 
the magnetic one. Notice that the relative value of $T^*$ and $T_{Neel}$  can 
slightly vary in different 122 compounds and different doping level depending on 
the band parameters. In particular for small $\d_e$, $T^{*}$ can be even lower 
than $T_{Neel}$ preventing the occurrence of a nematic phase.

\end{document}